\documentclass[11pt]{article}

\usepackage[a4paper,left=2.2cm,right=2.2cm,top=2.5cm,bottom=2.5cm]{geometry}

\usepackage[T1]{fontenc}
\usepackage{lmodern}

\usepackage{amsmath,amssymb,bm}
\usepackage{graphicx}
\usepackage{xcolor}
\usepackage{hyperref}

\begin{document}

\begin{center}
{\LARGE\bfseries Antisymmetric Mueller generator as the universal origin of geometric phase in classical polarization and quantum two-level systems\par}
\vspace{0.8em}

{\large Jose J. Gil\par}
\vspace{0.4em}

{\normalsize Independent Researcher, 29690 Casares Costa, Spain\par}
\vspace{0.3em}

\ttfamily {*ppgil@unizar.es\par}
\end{center}

\vspace{1.2em}

\begin{abstract}
We show that the antisymmetric Mueller generator provides a universal
algebraic kernel for geometric phase in classical polarization optics
and in quantum two-level systems. For any ideal retarder, the
antisymmetric $3\times 3$ block of its Mueller matrix---the
antisymmetric generator of the adjoint $\mathrm{SU}(2)$ action on the
Stokes vector---encodes the angular-velocity vector that drives the
tangential motion on the Poincar\'e sphere and fully determines the
Pancharatnam--Berry phase, while the symmetric block is geometrically
neutral. The same antisymmetric generator governs the evolution of
pure qubit states on the Bloch sphere. This unified viewpoint yields
operational criteria to identify and control geometric-phase
contributions from measured Mueller matrices and from qubit process
tomography.
\end{abstract}


\section{Introduction}

Geometric phase is a ubiquitous concept in modern physics, with
manifestations ranging from optics to quantum mechanics and condensed
matter~\cite{CohenNatRevPhys2019,Berry1984}. In classical polarization optics, the Pancharatnam--Berry (PB) phase arises
whenever the polarization state traces a closed contour on the Poincar\'e
sphere~\cite{Pancharatnam1956,Bhandari1997}. This effect underlies a wide
variety of polarization-based geometric-phase elements, from geometric-phase
lenses to wavefront-shaping metasurfaces, whose operation is routinely
described and characterized within the Mueller-matrix framework and related
polarimetric formalisms~\cite{RouxGPLens,ArteagaGP,ArteagaPiPhase}.

Several algebraic formulations have been developed to evaluate the PB
phase introduced by general polarization devices described by Jones
matrices, including systems with nonorthogonal eigenpolarizations~\cite{GutierrezVegaOL2011,LopezMagoOL2017}.
These approaches focus on the overall geometric phase associated with
a given optical system, typically in terms of its Jones representation
and the associated paths on the Poincar\'e sphere.

In quantum mechanics, Berry showed that a system undergoing adiabatic
and cyclic evolution acquires a geometric phase that depends only on
the path in parameter space~\cite{Berry1984}. This concept was
subsequently generalized to nonadiabatic and noncyclic evolutions by
Aharonov and Anandan~\cite{AharonovAnandan1987} and by Samuel and
Bhandari~\cite{SamuelBhandari1988}. For a two-level system, pure
states can be represented by points on the Bloch sphere, and the
associated geometric phases are again determined by solid angles on
that sphere. For polarized light, a comprehensive review of geometric
and topological phase phenomena is given in Ref.~\cite{Bhandari1997}. Further generalizations to mixed quantum states in an interferometric
setting were developed by Sj\"oqvist \textit{et al.}~\cite{Sjoeqvist2000}.

Although these developments are often presented in different formalisms (Jones and Mueller calculus in classical polarization, and Hilbert-space or SU(2)/Bloch-sphere representations in quantum mechanics) they share the same geometric origin: a curve on a space of states that can be identified with the two-sphere $S^2$. In both cases, pure states correspond to points on the Poincaré or Bloch sphere, and physically relevant evolutions are represented by rotations in $\mathrm{SO}(3)$ induced by the adjoint action of $\mathrm{SU}(2)$.

Beyond reproducing standard formulas for geometric phases in these familiar settings, the aim of this work is to present an explicitly algebraic and operational reformulation in which a single antisymmetric adjoint generator is identified as the common carrier of geometric-phase information in both regimes, and can be accessed directly from experimentally measured quantities

A practical point is that a Mueller (or adjoint) matrix describes an one-step state transformation, whereas a geometric phase is associated with a \emph{path} on the Poincar\'e/Bloch sphere. The connection is local: the antisymmetric block yields the angular-velocity generator that drives the tangential motion of the Stokes/Bloch vector. When a device parameter varies continuously (e.g., along propagation in a birefringent medium or under rotation of a wave plate), or when a sequence of such transformations is applied, the state traces a curve; the accumulated geometric phase then follows from the solid angle enclosed by that curve. In this sense, identifying $A$ is a differential statement: it fixes the instantaneous tangential velocity and thereby the geometric phase accumulated along any path.

In this work we make this correspondence algebraically explicit and
operational. We identify the antisymmetric Mueller generator, namely
the antisymmetric part of the adjoint $\mathrm{SU}(2)$ action on the
Stokes vector, as the universal kernel for geometric phase in both
classical polarization optics and quantum two-level systems. On the
classical side, we show that for any ideal retarder the antisymmetric
$3\times 3$ block of its Mueller matrix completely encodes the angular-velocity pseudovector that drives the tangential component of
the Stokes-vector motion and therefore fully determines the
Pancharatnam--Berry phase, while the complementary symmetric block is
geometrically neutral. On the quantum side, we show that the same
antisymmetric generator arises in the adjoint action of
$\mathrm{SU}(2)$ unitaries on the Bloch sphere and fully determines
the geometric phase of pure qubit states, independently of adiabaticity
or cyclicity. The role of the antisymmetric generator is again to encode the instantaneous angular velocity of the Bloch vector and, more specifically, the tangential component of this velocity on the Bloch sphere, which is the only part that contributes to the geometric phase.

These results identify a single algebraic structure underlying PB phases in classical polarization and geometric phases in quantum two-level systems: the antisymmetric part of the adjoint $\mathrm{SU}(2)$ generator. Beyond providing a unified conceptual framework, this viewpoint is
operational. In classical experiments, the antisymmetric Mueller
generator can be extracted directly from a measured Mueller matrix of
an ideal retarder. In quantum experiments, the same antisymmetric
generator can be reconstructed from process tomography through the
adjoint map acting on the Bloch vector. In both cases this yields a
practical criterion for isolating geometric-phase contributions and
for designing or diagnosing transformations where geometric phases
are controlled or suppressed.

The remainder of the paper is organized as follows. In Sec.~2 we analyze the Mueller matrix of an arbitrary ideal retarder and show that its antisymmetric block is the unique generator of geometric phase on the Poincaré sphere. In Sec.~3 we connect this antisymmetric generator with the tangential angular velocity of the Stokes vector and with the solid angle enclosed by its trajectory. Section~4 extends the formalism to quantum two-level systems by studying the adjoint action of $\mathrm{SU}(2)$ on the Bloch sphere. In Sec.~5 we present illustrative classical and quantum examples that highlight the operational content of the theory. Section~6 discusses possible extensions to more general quantum channels and higher-dimensional systems, and Sec.~7 summarizes our conclusions.

\section{Classical polarization: Mueller retarders and antisymmetric generator}
\label{sec:classical}

We first consider the classical case of an ideal retarder acting on a fully polarized beam. Here ``ideal retarder'' means a lossless, nondepolarizing, nondiattenuating element that introduces a relative phase delay between two orthogonal eigenpolarizations, without changing total intensity or degree of polarization. The retarder is described in the Jones formalism by a unitary $2\times 2$ operator of the form~\cite{Whitney1971}
\begin{equation}
  U = \exp\!\left[-\frac{i}{2}\,\delta\,\mathbf{n}\cdot\boldsymbol{\sigma}\right],
  \label{eq:jones-retarder}
\end{equation}
where $\delta$ is the retardance, $\mathbf{n}$ is a real unit vector that specifies the eigenpolarizations on the Poincaré sphere, and $\boldsymbol{\sigma} = (\sigma_1,\sigma_2,\sigma_3)$ is the Pauli vector,
i.e., the vector of Pauli matrices. The pair $(\delta,\mathbf{n})$ provides a complete parametrization of pure retarders: $\delta$ determines the rotation angle on the Poincaré sphere and $\mathbf{n}$ the rotation axis.

The Pauli-algebraic approach to polarization was developed by Whitney in her early work on Pauli operators~\cite{Whitney1971}. A systematic development
of a vectorial, pure operatorial Pauli-algebraic formalism for device
and state operators has been given by Tudor
\cite{TudorOptik2010,TudorApplOpt2012}, providing an alternative but
closely related framework to the standard Jones and Mueller calculi for deterministic (also called nondepolarizing) systems.

The corresponding action on the Stokes vector is given by a real $3\times 3$ rotation matrix $R \in \mathrm{SO}(3)$ acting as
\begin{equation}
  \mathbf{s}_{\mathrm{out}} = R\,\mathbf{s}_{\mathrm{in}},
\end{equation}
where $\mathbf{s} = (S_1,S_2,S_3)^{\mathsf T}$ denotes the normalized Stokes vector of a fully polarized state. The matrix $R$ is the adjoint $\mathrm{SO}(3)$ representation of the $\mathrm{SU}(2)$ operator $U$ and takes the standard Rodrigues form
(see, e.g., Refs.~\cite{GoldsteinCM,SakuraiNapolitano})
\begin{equation}
  R = \cos\delta\,\mathbb{I}_3
      + (1 - \cos\delta)\,\mathbf{n}\mathbf{n}^{\mathsf T}
      + \sin\delta\,[\mathbf{n}]_\times,
  \label{eq:R-decomposition}
\end{equation}
where $\mathbb{I}_3$ is the $3\times 3$ identity matrix,
$\mathbf{n}\mathbf{n}^{\mathsf T}$ is the rank-one projector onto $\mathbf{n}$, and $[\mathbf{n}]_\times$ is the cross-product matrix
\begin{equation}
  [\mathbf{n}]_\times =
  \begin{pmatrix}
    0      & -n_3  &  n_2 \\
    n_3    &  0    & -n_1 \\
    -n_2   &  n_1  &  0
  \end{pmatrix},
  \label{eq:cross-product-matrix}
\end{equation}
which satisfies $[\mathbf{n}]_\times\,\mathbf{v} = \mathbf{n}\times\mathbf{v}$ for any vector $\mathbf{v}$.

In the Mueller-matrix formalism, a pure (or ideal) retarder is represented by a $4\times 4$ Mueller matrix~\cite{GilOssikovski}:
\begin{equation}
  M =
  \begin{pmatrix}
    1 & \mathbf{0}^{\mathsf T} \\
    \mathbf{0} & R
  \end{pmatrix},
  \label{eq:mueller-retarder}
\end{equation}
where the upper-left unit element $m_{00} = 1$ reflects no attenuation and the $3\times 3$ submatrix $R$ acts on the reduced Stokes vector. Thus, for ideal retarders, all polarimetric structure relevant to the geometric phase is contained in $R$.

To identify the geometric generator, we decompose $R$ into its symmetric and antisymmetric parts,
\begin{equation}
  R = S + A, \qquad
  S = \frac{1}{2}\left(R + R^{\mathsf T}\right), \quad
  A = \frac{1}{2}\left(R - R^{\mathsf T}\right),
  \label{eq:S-A-decomposition}
\end{equation}
with $S^{\mathsf T} = S$ and $A^{\mathsf T} = -A$. Using Eq.~\eqref{eq:R-decomposition} and the antisymmetry of $[\mathbf{n}]_\times$, a straightforward calculation gives
\begin{equation}
  S = \cos\delta\,\mathbb{I}_3
      + (1 - \cos\delta)\,\mathbf{n}\mathbf{n}^{\mathsf T},
  \label{eq:symmetric-part}
\end{equation}
\begin{equation}
  A = \sin\delta\,[\mathbf{n}]_\times.
  \label{eq:antisymmetric-part}
\end{equation}
The symmetric part $S$ thus depends only on $\delta$ and on the projector along $\mathbf{n}$, whereas the antisymmetric part $A$ is proportional to the cross-product matrix $[\mathbf{n}]_\times$ with coefficient $\sin\delta$.

At the infinitesimal level relevant to geometric phase, only the antisymmetric part acts as a generator of motion on the sphere (via a cross product with an angular-velocity vector), whereas the symmetric part does not generate tangential transport and therefore cannot contribute to the swept solid angle.

From a geometric viewpoint, the antisymmetric matrix $A =
\sin\delta\,[\mathbf{n}]_\times$ can be identified with an axial
vector $\boldsymbol{\Omega} = \sin\delta\,\mathbf{n}$ whose direction
and sign fix the sense of rotation of Stokes vectors on the
Poincar\'e sphere.
Reversing $\boldsymbol{\Omega}$ reverses the circulation of any
trajectory and flips the sign of the corresponding geometric phase.
In this sense, the antisymmetric adjoint generator carries the chirality of the evolution on the sphere,
whereas the symmetric part $S$ is achiral and geometrically neutral.

Equations~\eqref{eq:symmetric-part} and~\eqref{eq:antisymmetric-part} show that the action of a pure retarder on the Poincaré sphere splits into two algebraically distinct components: a symmetric mapping that reshapes the distribution of Stokes vectors relative to the eigenaxis $\mathbf{n}$, and a rotational component governed by the antisymmetric block $A$. As will be discussed in Sec.~\ref{sec:geom_phase}, it is this rotational component that contains the geometric-phase information, whereas the symmetric part affects the mapping of Stokes vectors without contributing to the accumulated phase.

For any ideal retarder the antisymmetric $3\times 3$ block of the Mueller
matrix is therefore
\begin{equation}
  A = \sin\delta\,[\mathbf{n}]_\times,
  \label{eq:A-final}
\end{equation}
and this matrix will be identified in the following sections as the universal
classical kernel of geometric phase.

\begin{figure}[t]
  \centering
  \includegraphics[width=0.45\textwidth]{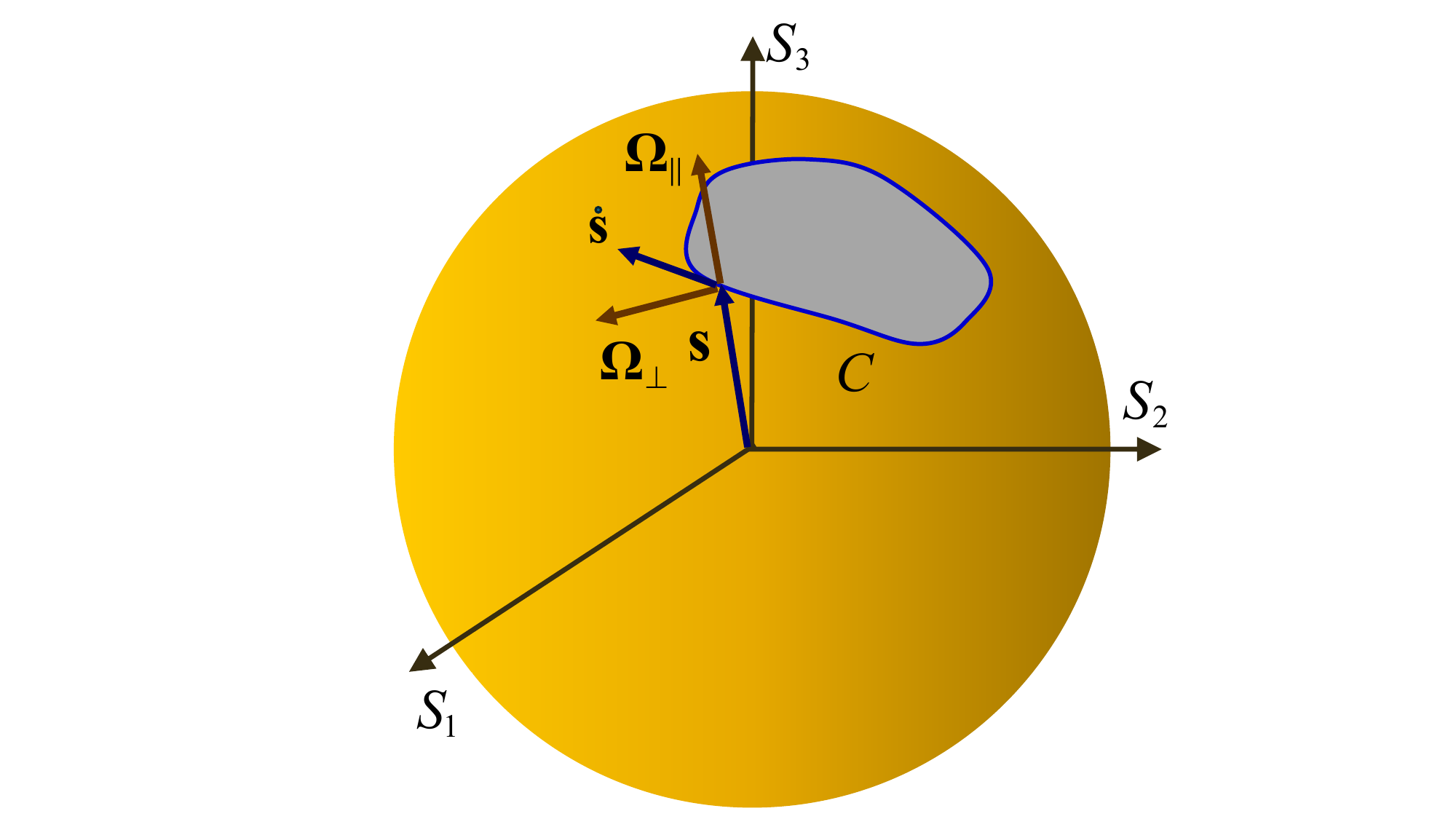}
  \caption{Representative closed trajectory on the Poincar\'e/Bloch sphere
  in $(S_1,S_2,S_3)$ coordinates. The closed curve and its shaded interior
  represent the solid angle enclosed by the Stokes/Bloch vector.
  The instantaneous tangential component of the angular velocity,
  $\boldsymbol{\Omega}_\perp$, lies in the tangent plane to the sphere
  and is generated exclusively by the antisymmetric adjoint operator
  $A = \sin\delta\,[\mathbf{n}]_\times$ for classical pure retarders and
  by $A_U = \sin\delta\,[\mathbf{n}]_\times$ for quantum two-level
  unitaries. This common antisymmetric generator fully determines
  the geometric phase. The actual velocity of the Stokes/Bloch vector
  along the trajectory is $\dot{\mathbf{s}} =
  \boldsymbol{\Omega}_\perp \times \mathbf{s}$.}
  \label{fig:sphere}
\end{figure}

\section{Geometric phase and tangential angular velocity on the Poincar\'e sphere}
\label{sec:geom_phase}

The decomposition \eqref{eq:S-A-decomposition}--\eqref{eq:antisymmetric-part}
shows that, although the complete polarimetric action of an ideal retarder on
the Poincar\'e sphere is described by the rotation $R$, the geometric-phase
behavior is governed by an antisymmetric matrix $A=[\boldsymbol{\Omega}]_\times$ with
\begin{equation}
  \boldsymbol{\Omega} = \sin\delta\,\mathbf{n},
  \label{eq:Omega-def}
\end{equation}
where we have introduced the angular-velocity vector $\boldsymbol{\Omega}$
associated with the rotation around $\mathbf{n}$ in the usual sense of rigid-body
kinematics on the sphere. Indeed, for any Stokes vector $\mathbf{s}$ we have
\begin{equation}
  A\,\mathbf{s} = [\boldsymbol{\Omega}]_\times\,\mathbf{s}
  = \boldsymbol{\Omega}\times\mathbf{s},
\end{equation}
so that the infinitesimal action of the retarder on $\mathbf{s}$ can be written
in the kinematic form
\begin{equation}
  \frac{d\mathbf{s}}{dt} = \boldsymbol{\Omega}(t)\times\mathbf{s}(t),
  \label{eq:kinematic-eq}
\end{equation}
where $t$ denotes any convenient parameter along the evolution (for instance,
the physical time, the propagation distance inside the medium, or a control
parameter such as the rotation angle of a wave plate). In this sense
$\boldsymbol{\Omega}(t)$ represents the instantaneous angular velocity of
$\mathbf{s}(t)$ on the Poincar\'e sphere, and Eq.~\eqref{eq:kinematic-eq} is
the standard equation of motion for a unit vector driven by such an angular
velocity.

More generally, an arbitrary retarder can be viewed as a concatenation
of elementary retarders with piecewise constant parameters
$(\delta,\mathbf{n})$ or, in the continuum limit, as generated by a
smooth angular-velocity field $\boldsymbol{\Omega}(t)$ along the
evolution parameter. The net action is always equivalent to a single
rotation on the sphere with some effective generator
$[\boldsymbol{\Omega}_{\mathrm{eff}}]_\times$, although
$\boldsymbol{\Omega}_{\mathrm{eff}}$ is in general a nontrivial
function of the full history of $\boldsymbol{\Omega}(t)$ due to the
noncommutativity of rotations. A physically relevant example of such
a continuum limit is provided by an optical fiber whose local
birefringence and eigenpolarizations vary smoothly along the
propagation direction, so that the Stokes vector experiences a
continuously varying angular velocity on the Poincar\'e sphere.

The geometric phase associated with the evolution of a fully polarized state can be understood as a property of the trajectory $\mathbf{s}(t)$ on the Poincar\'e sphere, rather than as a property of the underlying Jones vector itself. For a closed loop $C$ traced by $\mathbf{s}(t)$, the Pancharatnam--Berry (PB) phase is \cite{Pancharatnam1956,Berry1984,SamuelBhandari1988,Bhandari1997}
\begin{equation}
  \gamma_{\mathrm{geom}}(C) = -\frac{1}{2}\,\Omega_{\mathrm{solid}}(C),
  \label{eq:gamma-solid-angle}
\end{equation}
where $\Omega_{\mathrm{solid}}(C)$ is the oriented solid angle subtended by $C$ at the center of the sphere (see Fig.~\ref{fig:sphere}). The expression \eqref{eq:gamma-solid-angle} can be derived from the
Pancharatnam connection
\begin{equation}
\gamma = \arg\langle \psi(0)|\psi(T)\rangle - \gamma_{\rm dyn},
\end{equation}
where $\arg\langle \psi(0)|\psi(T)\rangle$ is the total phase difference
between the initial and final states and $\gamma_{\rm dyn}$ is the dynamical
phase. The overlap between initial and final states on the sphere is governed
by the solid angle enclosed by the trajectory. For a comprehensive geometric
treatment see~\cite{Bhandari1997,SamuelBhandari1988}.

This expression can be written in differential form as a line integral over the curve or, equivalently, as a surface integral over any surface $\Sigma$ bounded by $C$.

Equivalently, one may view Eq.~\eqref{eq:gamma-solid-angle} as a direct consequence of the Gauss--Bonnet theorem on the sphere~\cite{BhandariPLA1989}: for a closed loop completed by geodesic segments (as in Pancharatnam's geodesic closure), the enclosed oriented area is tied to the integrated geodesic curvature and fixes the geometric phase.

\begin{figure}[t]
  \centering
  \includegraphics[width=0.45\textwidth]{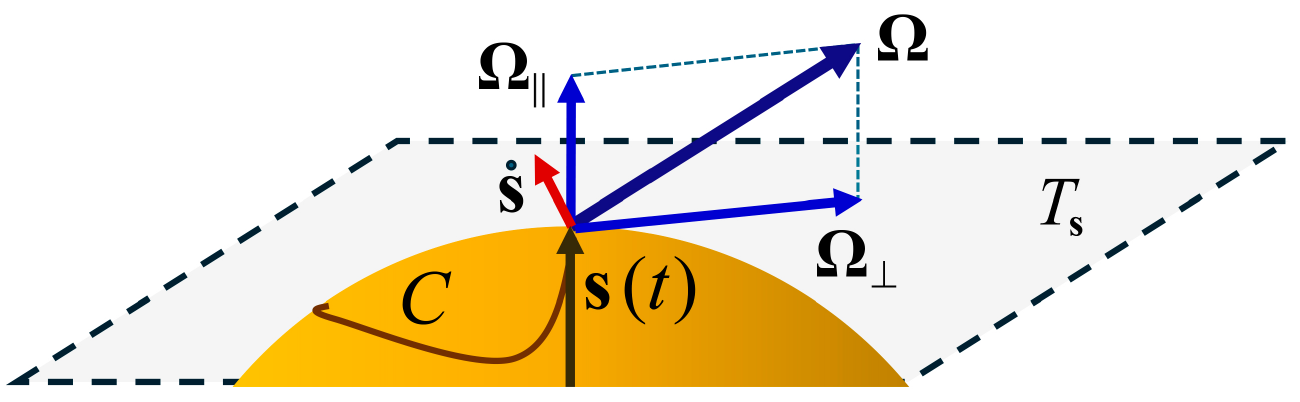}
 \caption{Local geometry of the Stokes/Bloch vector and the antisymmetric
adjoint generator at a point $\mathbf{s}$ on the Poincar\'e/Bloch sphere.
For visual clarity the point is shown at the north pole of the sphere,
but the construction is local and applies to any point.
The unit vector $\mathbf{s}$ is normal to the sphere and to the tangent plane $T_{\mathbf{s}}$
at that point. The instantaneous angular velocity
$\boldsymbol{\Omega}$ is decomposed into a component
$\boldsymbol{\Omega}_\parallel$ parallel to $\mathbf{s}$ and a component
$\boldsymbol{\Omega}_\perp$ that lies in the tangent plane.
The velocity of the Stokes/Bloch vector is
$\dot{\mathbf{s}}=\boldsymbol{\Omega}_\perp\times\mathbf{s}$; it is therefore
tangent to the sphere and orthogonal to $\boldsymbol{\Omega}_\perp$ within the
tangent plane. Reversing $\boldsymbol{\Omega}_\perp$ reverses the circulation
of the trajectory and the sign of the geometric phase.}
  \label{fig:local}
\end{figure}

To make the connection with Eq.~\eqref{eq:kinematic-eq}, we recall that the solid angle can be expressed as a surface integral of the Berry curvature on the sphere. Let $\mathbf{s}(\theta,\phi)$ denote the unit vector corresponding to spherical angles $(\theta,\phi)$, with $0\leq\theta\leq\pi$ and $0\leq\phi<2\pi$, and let $\Sigma$ be a surface on $S^2$ with boundary $C$. Then
\begin{equation}
  \Omega_{\mathrm{solid}}(C)
  = \iint_{\Sigma} \mathcal{F}(\theta,\phi)\,d\theta\,d\phi,
\end{equation}
where the Berry curvature two-form on the sphere is (see, e.g., Refs.~\cite{Berry1984,Nakahara})
\begin{equation}
  \mathcal{F}(\theta,\phi)
  = \mathbf{s}\cdot\left(\frac{\partial\mathbf{s}}{\partial\theta}
  \times \frac{\partial\mathbf{s}}{\partial\phi}\right)
  \label{eq:berry-curvature-sphere}
\end{equation}
and reduces to $\mathcal{F} = \sin\theta$ in the usual parametrization. Equation~\eqref{eq:gamma-solid-angle} then becomes
\begin{equation}
  \gamma_{\mathrm{geom}}(C)
  = -\frac{1}{2}\iint_{\Sigma}
  \mathbf{s}\cdot\left(\frac{\partial\mathbf{s}}{\partial\theta}
  \times \frac{\partial\mathbf{s}}{\partial\phi}\right)
  d\theta\,d\phi.
  \label{eq:gamma-surface}
\end{equation}

On the other hand, the kinematic equation~\eqref{eq:kinematic-eq} implies that the instantaneous motion of $\mathbf{s}(t)$ on the sphere is characterized by the angular velocity $\boldsymbol{\Omega}(t)$. We can decompose $\boldsymbol{\Omega}(t)$ into components parallel and orthogonal to $\mathbf{s}(t)$,
\begin{equation}
  \boldsymbol{\Omega}(t)
  = \boldsymbol{\Omega}_\parallel(t) + \boldsymbol{\Omega}_\perp(t),
  \qquad
  \boldsymbol{\Omega}_\parallel
  = (\boldsymbol{\Omega}\cdot\mathbf{s})\,\mathbf{s},
  \quad
  \boldsymbol{\Omega}_\perp
  = \boldsymbol{\Omega} - \boldsymbol{\Omega}_\parallel,
  \label{eq:Omega-decomp}
\end{equation}
where $\boldsymbol{\Omega}_\parallel$ is the component parallel to $\mathbf{s}$ and $\boldsymbol{\Omega}_\perp$ lies in the tangent plane of the sphere at $\mathbf{s}$. 

The local geometry of these quantities is illustrated in Fig.~\ref{fig:local}. 
At a given time $t$, the unit Stokes/Bloch vector $\mathbf{s}(t)$ is normal to the
sphere and therefore normal to the tangent plane $T_{\mathbf{s}}$ at that
point.

The kinematic equation $\dot{\mathbf{s}}=\boldsymbol{\Omega}\times\mathbf{s}$
shows that the velocity $\dot{\mathbf{s}}(t)$ of the Stokes/Bloch vector is also
tangent to the sphere and is in fact determined solely by the tangential
component of the angular velocity,
$\dot{\mathbf{s}}(t)=\boldsymbol{\Omega}_\perp(t)\times\mathbf{s}(t)$, since
$\boldsymbol{\Omega}_\parallel(t)\times\mathbf{s}(t)=\mathbf{0}$. As a
consequence, both $\boldsymbol{\Omega}_\perp(t)$ and $\dot{\mathbf{s}}(t)$ lie
in the tangent plane and are mutually orthogonal there. Reversing $\boldsymbol{\Omega}_\perp$ reverses the circulation of the trajectory
on the sphere and flips the sign of the associated geometric phase. This makes
explicit that the tangential component of the antisymmetric adjoint generator
encodes not only the magnitude of the instantaneous angular velocity, but also
the handedness of the evolution on the Poincar\'e/Bloch sphere.

From a more intuitive viewpoint, the component of the angular velocity that is parallel to the Stokes/Bloch vector merely changes an overall phase convention and leaves the point fixed on the sphere, so it cannot contribute to the swept solid angle, whereas only the component that lies in the tangent plane actually displaces the state along a curve and is therefore responsible for the geometric phase.

Equation~\eqref{eq:kinematic-eq} then shows that the motion of $\mathbf{s}(t)$
on the sphere is driven entirely by $\boldsymbol{\Omega}_\perp$, since
\begin{equation}
  \boldsymbol{\Omega}_\parallel(t)\times\mathbf{s}(t) = \mathbf{0},
  \qquad
  \boldsymbol{\Omega}_\perp(t)\times\mathbf{s}(t)
  = \boldsymbol{\Omega}(t)\times\mathbf{s}(t).
\end{equation}
Thus $\boldsymbol{\Omega}_\parallel$ is collinear with $\mathbf{s}$ and does not
induce any motion on the sphere, whereas $\boldsymbol{\Omega}_\perp$ fully
drives the evolution of $\mathbf{s}(t)$. In particular, the component of
$\boldsymbol{\Omega}_\perp$ along the local normal to the osculating plane
determines the geodesic curvature of the trajectory and hence the sign and
magnitude of the geometric phase.

The PB phase can be expressed in terms of $\boldsymbol{\Omega}_\perp(t)$ by noting that the solid angle enclosed by the trajectory can be written as an integral over time of the oriented area swept by $\mathbf{s}(t)$ on the sphere. A convenient way to express this is to parametrize the curve by $t$ and write
\begin{equation}
  \Omega_{\mathrm{solid}}(C)
  = \int_0^T dt\,\kappa_{\mathrm{g}}(t)\,v(t),
  \label{eq:solid-angle-kinematic}
\end{equation}
where $v(t) = |\dot{\mathbf{s}}(t)|$ is the speed of $\mathbf{s}(t)$ along the curve (with respect to the standard metric on $S^2$), and $\kappa_{\mathrm{g}}(t)$ is the geodesic curvature of the trajectory. For unit-speed motion one has the classical relation $\kappa_{\mathrm{g}} = \boldsymbol{\Omega}_\perp\cdot\mathbf{u}_\mathrm{n}$, where $\mathbf{u}_\mathrm{n}$ is the normal to the osculating plane, and the sign of $\kappa_{\mathrm{g}}$ captures the orientation of the curve. Thus the geometric phase can be written in the kinematic form
\begin{equation}
  \gamma_{\mathrm{geom}}(C)
  = -\frac{1}{2}\int_0^T dt\,
  \kappa_{\mathrm{g}}(t)\,v(t),
  \label{eq:gamma-kg}
\end{equation}
which shows explicitly that $\gamma_{\mathrm{geom}}$ depends only on the tangential component $\boldsymbol{\Omega}_\perp(t)$ of the angular velocity.

Taken together, Eqs.~\eqref{eq:Omega-def}--\eqref{eq:gamma-kg} show that, for ideal retarders, the evolution relevant to geometric phase is governed entirely by the antisymmetric part of the Mueller block. The symmetric contribution does not appear in the kinematic description of the motion on the sphere and therefore plays no role in determining the solid angle or the geometric phase. In other words, the antisymmetric $3\times 3$ block $A$ of the Mueller matrix
contains the complete geometric-phase content of the retarder: it specifies the
instantaneous angular velocity $\boldsymbol{\Omega}$ and, in particular, its
tangential component $\boldsymbol{\Omega}_\perp$ that controls the PB phase
through Eq.~\eqref{eq:gamma-kg}.

Although the discussion above is phrased in terms of classical polarization, the kinematic equation~\eqref{eq:kinematic-eq} and the decomposition~\eqref{eq:Omega-decomp} are purely geometric statements about curves on the two-sphere. As such, they apply equally well to the evolution of Bloch vectors representing pure states of a quantum two-level system. In the next section we show that the adjoint action of any $\mathrm{SU}(2)$ unitary operator on the Bloch sphere is generated by an antisymmetric matrix of the same form \eqref{eq:antisymmetric-part}, and that the corresponding geometric phase is again fully determined by this antisymmetric generator.

\section{Quantum two-level systems: adjoint SU(2) action and Bloch-sphere geometric phase}
\label{sec:quantum}

We now turn to quantum two-level systems and show that the same antisymmetric generator that appears in the Mueller description of pure retarders also governs geometric phases on the Bloch sphere.
Any qubit state can be represented by a Bloch vector $\mathbf{r}$,
with pure states lying on the unit sphere and mixed states in its
interior, and their dynamics is generated by $\mathrm{SU}(2)$
unitaries in the adjoint representation~\cite{SakuraiNapolitano,NielsenChuangBook}. For cyclic evolutions of pure states, the
Aharonov--Anandan phase~\cite{AharonovAnandan1987} can be written as
\begin{equation}
  \gamma_{\mathrm{geom}}(C) = -\frac{1}{2}\,\Omega_{\mathrm{solid}}(C),
\end{equation}
in direct analogy with the classical PB phase.

Working in the Bloch-vector representation is automatically gauge invariant, because the dynamics is encoded in density matrices and real three-dimensional trajectories, so no explicit choice of phase for the underlying state vector is required; the geometric phase is directly associated with the geometry of the path on the Bloch sphere.

Consider a qubit with density matrix
\begin{equation}
  \rho = \frac{1}{2}\left(\mathbb{I}_2 + \mathbf{r}\cdot\boldsymbol{\sigma}\right),
  \label{eq:qubit-density}
\end{equation}
where $\mathbf{r}\in\mathbb{R}^3$ is the Bloch vector, $|\mathbf{r}| \leq 1$, and $\boldsymbol{\sigma}$ is the Pauli vector. Pure states correspond to $|\mathbf{r}|=1$, so that $\mathbf{r}$ lies on the Bloch sphere. 

Let $U(t)\in\mathrm{SU}(2)$ be a unitary evolution operator acting on the qubit, with $U(0) = \mathbb{I}_2$. The density matrix evolves as
\begin{equation}
  \rho(t) = U(t)\,\rho(0)\,U^\dagger(t),
\end{equation}
which induces an adjoint action on the Bloch vector,
\begin{equation}
  \mathbf{r}(t) = R_U(t)\,\mathbf{r}(0),
  \label{eq:bloch-adjoint}
\end{equation}
where $R_U(t)\in\mathrm{SO}(3)$ is a real rotation matrix. The mapping $U\mapsto R_U$ is the adjoint representation of $\mathrm{SU}(2)$ on $\mathbb{R}^3$ and is formally identical to the mapping that relates the Jones operator \eqref{eq:jones-retarder} to the Poincar\'e-sphere rotation \eqref{eq:R-decomposition}.

Any $U(t)\in\mathrm{SU}(2)$ can be written as
\begin{equation}
  U(t) = \exp\!\left[-\frac{i}{2}\,\delta(t)\,\mathbf{n}(t)\cdot\boldsymbol{\sigma}\right],
  \label{eq:U-param}
\end{equation}
where $\delta(t)$ is a time-dependent rotation angle and $\mathbf{n}(t)$ is a unit vector specifying the instantaneous rotation axis in Bloch space. Inserting Eq.~\eqref{eq:U-param} into the adjoint map and using standard $\mathrm{SU}(2)\to\mathrm{SO}(3)$ identities~\cite{SakuraiNapolitano}, one finds that $R_U(t)$ has exactly the same structure as Eq.~\eqref{eq:R-decomposition},
\begin{equation}
  R_U(t)
  = \cos\delta(t)\,\mathbb{I}_3
      + \bigl[1 - \cos\delta(t)\bigr]\,\mathbf{n}(t)\mathbf{n}^{\mathsf T}(t)
      + \sin\delta(t)\,[\mathbf{n}(t)]_\times.
  \label{eq:RU-decomposition}
\end{equation}
The corresponding antisymmetric and symmetric parts are
\begin{equation}
  S_U(t) = \cos\delta(t)\,\mathbb{I}_3
      + \bigl[1 - \cos\delta(t)\bigr]\,\mathbf{n}(t)\mathbf{n}^{\mathsf T}(t),
  \label{eq:SU-part}
\end{equation}
\begin{equation}
  A_U(t) = \sin\delta(t)\,[\mathbf{n}(t)]_\times.
  \label{eq:AU-part}
\end{equation}
Thus, the adjoint action of $U(t)$ on the Bloch vector is generated by an antisymmetric matrix of the form
\begin{equation}
  A_U(t) = [\boldsymbol{\Omega}(t)]_\times,
  \qquad
  \boldsymbol{\Omega}(t) = \sin\delta(t)\,\mathbf{n}(t),
  \label{eq:Omega-U-def}
\end{equation}
which leads to the kinematic equation
\begin{equation}
  \frac{d\mathbf{r}(t)}{dt}
  = \boldsymbol{\Omega}(t)\times\mathbf{r}(t),
  \label{eq:bloch-kinematic}
\end{equation}
identical in form to Eq.~\eqref{eq:kinematic-eq}. In this picture
$\mathbf{r}(t)$ is a unit vector on the Bloch sphere and
$\boldsymbol{\Omega}(t)$ plays the role of its instantaneous angular
velocity on the sphere, now determined by the underlying
$\mathrm{SU}(2)$ Hamiltonian, in direct analogy with the Stokes-vector
evolution discussed in Sec.~\ref{sec:geom_phase}. 

As in the classical retarder case, the overall unitary evolution over
a finite interval can always be represented by a single effective
rotation on the Bloch sphere with generator
$[\boldsymbol{\Omega}_{\mathrm{eff}}]_\times$, although
$\boldsymbol{\Omega}_{\mathrm{eff}}$ is in general a nontrivial
functional of the full history of $\boldsymbol{\Omega}(t)$ because of
the noncommutativity of $\mathrm{SU}(2)$ rotations.

As in the classical
case, we can decompose $\boldsymbol{\Omega}(t)$ into components
parallel and orthogonal to $\mathbf{r}(t)$,
\begin{equation}
  \boldsymbol{\Omega}(t)
  = \boldsymbol{\Omega}_\parallel(t) + \boldsymbol{\Omega}_\perp(t),
\end{equation}
with $\boldsymbol{\Omega}_\parallel = (\boldsymbol{\Omega}\cdot\mathbf{r})\,\mathbf{r}$ and $\boldsymbol{\Omega}_\perp = \boldsymbol{\Omega} - \boldsymbol{\Omega}_\parallel$.

The geometric phase associated with the evolution of a pure qubit state can again be written as a property of the trajectory $\mathbf{r}(t)$ on the Bloch sphere. For a cyclic evolution where $\mathbf{r}(T) = \mathbf{r}(0)$, the Aharonov--Anandan geometric phase~\cite{AharonovAnandan1987} is given by
\begin{equation}
  \gamma_{\mathrm{geom}}(C)
  = -\frac{1}{2}\,\Omega_{\mathrm{solid}}(C),
  \label{eq:gamma-bloch-solid}
\end{equation}
where $C$ is the closed Bloch trajectory and $\Omega_{\mathrm{solid}}(C)$ is the corresponding solid angle. This expression holds independently of whether the evolution is adiabatic or nonadiabatic and reduces to Berry's original formula in the adiabatic case. It is precisely the Aharonov--Anandan geometric phase for cyclic evolutions of pure states and therefore provides a nonadiabatic extension of Berry's result within the same geometric framework.

By the same geometric arguments as in Sec.~\ref{sec:geom_phase}, the solid angle can be expressed in terms of the tangential component $\boldsymbol{\Omega}_\perp(t)$ of the angular velocity of $\mathbf{r}(t)$ on the sphere. In particular, one can write
\begin{equation}
  \gamma_{\mathrm{geom}}(C)
  = -\frac{1}{2}\int_0^T dt\,
  \kappa_{\mathrm{g}}(t)\,v(t),
  \label{eq:gamma-bloch-kg}
\end{equation}
where $v(t) = |\dot{\mathbf{r}}(t)|$ is the speed of the Bloch vector along $C$ and $\kappa_{\mathrm{g}}(t)$ is the geodesic curvature of the trajectory. The geodesic curvature depends only on the tangential component $\boldsymbol{\Omega}_\perp(t)$ of the angular velocity and vanishes for geodesic motion on the sphere. 
Equation~\eqref{eq:gamma-bloch-kg} shows that, for pure-state qubit evolutions, the geometric contribution to the phase is encoded in the antisymmetric part of the adjoint map through the tangential angular velocity of the Bloch vector. The remaining components of the transformation do not influence the enclosed solid angle and hence do not affect the geometric phase.

From the operational point of view, this means that the geometric-phase
content of a unitary qubit evolution can be extracted solely from the
antisymmetric part of the adjoint map $R_U(t)$. In experiments where the quantum process is characterized via qubit
process tomography, one reconstructs the completely positive
trace-preserving map $\mathcal{E}$ on qubit density matrices
$\rho$, i.e., the map $\rho \mapsto \mathcal{E}(\rho)$, and the
induced action on the Bloch vector,
\begin{equation}
  \mathbf{r}_{\mathrm{out}} = R_{\mathcal{E}}\,\mathbf{r}_{\mathrm{in}} + \mathbf{t},
\end{equation}
where $R_{\mathcal{E}}$ is a real $3\times 3$ matrix and $\mathbf{t}$ is a translation vector.

\section{Illustrative examples}
\label{sec:examples}

In this section we present two simple examples that illustrate the role of the antisymmetric adjoint generator as the unique source of geometric phase. The first is a classical polarization experiment based on a rotating wave plate, and the second is a quantum spin-$1/2$ system in a cyclically varying magnetic field.
\subsection{Classical example: rotating wave plate on the Poincar\'e sphere}

Consider a (lossless) ideal wave plate with retardance $\delta$ and fast axis defined by a unit vector $\mathbf{n}(\varphi)$ on the Poincar\'e sphere, where
$\varphi$ is the azimuth of the fast-axis eigenpolarization in the laboratory
frame. For definiteness, we take the fast axis to lie initially in the equatorial plane and to rotate around the $S_3$ axis, so that
\begin{equation}
  \mathbf{n}(\varphi) =
  \bigl(\cos 2\varphi,\;\sin 2\varphi,\;0\bigr)^{\mathsf T}.
  \label{eq:n-varphi}
\end{equation}
The Jones operator of the plate is
\begin{equation}
  U(\varphi) = \exp\!\left[-\frac{i}{2}\,\delta\,\mathbf{n}(\varphi)\cdot\boldsymbol{\sigma}\right],
\end{equation}
and the corresponding Poincar\'e-sphere rotation is given by Eq.~\eqref{eq:R-decomposition} with $\mathbf{n}\to\mathbf{n}(\varphi)$. The antisymmetric generator is
\begin{equation}
  A(\varphi) = \sin\delta\,[\mathbf{n}(\varphi)]_\times
  = [\boldsymbol{\Omega}(\varphi)]_\times,
  \qquad
  \boldsymbol{\Omega}(\varphi) = \sin\delta\,\mathbf{n}(\varphi).
\end{equation}

Let the input state be a linear polarization along the $S_1$ axis,
\begin{equation}
  \mathbf{s}_{\mathrm{in}} = (1,0,0)^{\mathsf T}.
\end{equation}
As the plate rotates quasistatically from $\varphi=0$ to $\varphi=2\pi$ the Stokes vector $\mathbf{s}(\varphi)$ traces a closed curve $C$ on the Poincar\'e sphere. 

For a quarter-wave plate ($\delta=\pi/2$) and a fixed input state, the resulting trajectory is a closed, generally nonplanar loop on the Poincar\'e sphere. In the standard configuration of a fixed linear input polarization and a rotating linear retarder, this loop is the well-known three-dimensional figure-eight contour, given by the line of intersection of the Poincar\'e sphere with a right-circular cylinder whose axis is parallel to the $S_3$ direction \cite{AzzamJOSAA2000}. For a half-wave plate ($\delta=\pi$) and the same fixed linear input, the trajectory remains confined to the equatorial plane (corresponding to linear polarization states), so that no solid angle is enclosed and the geometric phase vanishes.

 In all cases, the instantaneous motion of $\mathbf{s}(\varphi)$ is governed by
\begin{equation}
  \frac{d\mathbf{s}(\varphi)}{d\varphi}
  = \boldsymbol{\Omega}(\varphi)\times\mathbf{s}(\varphi),
\end{equation}
and the geometric phase accumulated over a full $2\pi$ rotation is given by
\begin{equation}
  \gamma_{\mathrm{geom}}(C)
  = -\frac{1}{2}\,\Omega_{\mathrm{solid}}(C),
\end{equation}
where $\Omega_{\mathrm{solid}}(C)$ is the solid angle enclosed by the trajectory.

The explicit evaluation of $\Omega_{\mathrm{solid}}(C)$ depends on $\delta$ and on the initial state $\mathbf{s}_{\mathrm{in}}$, but its origin can be traced entirely to the antisymmetric generator $A(\varphi)$, since the symmetric part $S(\varphi)$ of the Mueller block does not enter the kinematic equation. For example, for a quarter-wave plate and a suitably chosen input state, the trajectory is a circle of colatitude $\theta$ on the sphere, and one obtains
\begin{equation}
  \gamma_{\mathrm{geom}}(C)
  = -\pi\bigl(1-\cos\theta\bigr),
\end{equation}
in agreement with the standard PB-phase result for such configurations. The present formalism shows that this phase can be read directly from $\boldsymbol{\Omega}(\varphi)$, which is determined by the antisymmetric part of the Mueller matrix of the rotating plate.

More generally, the optical retarding behavior of an arbitrary linear polarimetric transformation described by a Mueller matrix $M$ is completely captured by a pair of equivalent ideal retarders (the entrance and exit retarders), which encode the geometric phase associated with each pair of incident and emerging polarization states~\cite{GilSanJoseOssikovskiJOSAA2024}. 

\subsection{Quantum example: spin-1/2 particle in a cyclic magnetic field}

As a quantum counterpart, consider a spin-$1/2$ particle in a magnetic field of fixed magnitude $B$ whose direction precesses around the $z$ axis at a fixed polar angle $\theta$. Here the unit vector $\mathbf{n}(t)$ denotes the \emph{physical} field direction in real space (laboratory coordinates), and the same $\mathbf{n}(t)$ enters the Bloch-sphere dynamics because the Hamiltonian generates rotations of the Bloch vector about the instantaneous field axis. The Hamiltonian is
\begin{equation}
\begin{aligned}
  H(t) &= -\frac{\hbar \omega}{2}\,\mathbf{n}(t)\cdot\boldsymbol{\sigma},\\
  \mathbf{n}(t) &=
  \bigl(\sin\theta\cos\phi(t),\;
        \sin\theta\sin\phi(t),\;
        \cos\theta\bigr)^{\mathsf T}.
\end{aligned}
\label{eq:H-spin}
\end{equation}
where $\omega$ is the Larmor frequency and $\phi(t)$ increases monotonically from $0$ to $2\pi$ during a period $T$. The corresponding unitary evolution operator $U(t)$ is of the form \eqref{eq:U-param}, and its adjoint action on the Bloch vector is described by Eq.~\eqref{eq:RU-decomposition} with $\mathbf{n}\to\mathbf{n}(t)$.

If the system is prepared initially in the instantaneous ground state of $H(0)$ and the field is varied adiabatically, the Bloch vector $\mathbf{r}(t)$ follows $\mathbf{n}(t)$ and traces a cone of polar angle $\theta$ on the Bloch sphere. After one cycle the state returns to itself up to a phase, and the Berry phase is~\cite{Berry1984}
\begin{equation}
  \gamma_{\mathrm{geom}}(C)
  = -\pi\bigl(1-\cos\theta\bigr),
\end{equation}
which coincides with $-\frac{1}{2}$ times the solid angle $2\pi(1-\cos\theta)$ enclosed by the cone. From the viewpoint of the present formalism, this phase arises from the tangential component $\boldsymbol{\Omega}_\perp(t)$ of the angular velocity of $\mathbf{r}(t)$, where
\begin{equation}
  A_U(t) = [\boldsymbol{\Omega}(t)]_\times,
  \qquad \boldsymbol{\Omega}(t) = \sin\delta(t)\,\mathbf{n}(t),
\end{equation}
and in this setting the time-dependent rotation angle can be identified with the dynamical precession angle generated by the Hamiltonian, that is, with the time integral of the Larmor frequency for a field of fixed magnitude, so it plays the role of a rotation angle in Bloch space. The symmetric part $S_U(t)$ of the adjoint map does not contribute to $\gamma_{\mathrm{geom}}$.

For nonadiabatic evolutions where $\mathbf{r}(t)$ does not exactly follow $\mathbf{n}(t)$, the trajectory $C$ on the Bloch sphere changes but the geometric phase is still given by Eq.~\eqref{eq:gamma-bloch-kg}, with $\boldsymbol{\Omega}(t)$ determined by the antisymmetric generator $A_U(t)$. The explicit value of $\gamma_{\mathrm{geom}}$ now depends on the detailed dynamics, but its geometric origin and its dependence on $\boldsymbol{\Omega}_\perp(t)$ remain unchanged. This illustrates the fact that the antisymmetric adjoint generator provides a unified description of geometric phases in both adiabatic and nonadiabatic regimes.

\section{Operational aspects and possible extensions}
\label{sec:operational}

The identification of the antisymmetric adjoint generator as the universal kernel of geometric phase has direct operational implications in both classical polarization and quantum process characterization.

\subsection{Extraction from measured Mueller matrices}

In classical polarization optics, the Mueller matrix associated with the action of a linear system can be measured experimentally using standard polarimetric techniques. The procedure to extract the entrance and
exit (ideal) retarders of a general Mueller matrix is described in Ref.~\cite{GilSanJoseOssikovskiJOSAA2024}. For a given ideal retarder, the Mueller matrix has the block form \eqref{eq:mueller-retarder}, with a $3\times 3$ submatrix $R$ that must belong to $\mathrm{SO}(3)$. Once $R$ is known, its antisymmetric part
\begin{equation}
  A = \frac{1}{2}\left(R - R^{\mathsf T}\right)
\end{equation}
is directly obtained and can be written as $A = [\boldsymbol{\Omega}]_\times$. The vector $\boldsymbol{\Omega}$ specifies the instantaneous angular velocity of the Stokes vector under the action of the retarder and, as shown in Sec.~\ref{sec:geom_phase}, fully determines the geometric phase associated with any trajectory generated by the device. In realistic situations where the measured Mueller matrix deviates slightly from that of an ideal retarder due to weak diattenuation or depolarization, the antisymmetric part of the effective retarding block still provides a meaningful approximate geometric generator, capturing the dominant rotational contribution to the phase while the nonideal features mainly affect the symmetric and depolarizing components.

This provides a simple criterion for identifying and controlling geometric-phase effects in polarization experiments. For instance, if a composite optical system is designed such that the net antisymmetric generator $A$ vanishes over a given range of parameters, then its net geometric phase is suppressed, even if the symmetric part of the Mueller block induces nontrivial transformations of Stokes vectors. Conversely, by tailoring $A$ one can design devices that implement prescribed geometric phases with a given trajectory on the Poincar\'e sphere.

\subsection{Quantum process tomography and adjoint maps}

In quantum information experiments, the evolution of
a qubit is often characterized via quantum process tomography~\cite{NielsenChuangBook,ChuangNielsenQPT}.
In the Bloch representation, a general quantum channel
$\mathcal{E}$ acts as
\begin{equation}
  \rho \mapsto \mathcal{E}(\rho), \qquad
  \mathbf{r}_{\mathrm{out}} = R_{\mathcal{E}}\,\mathbf{r}_{\mathrm{in}} + \mathbf{t},
  \label{eq:affine-map}
\end{equation}
where $R_{\mathcal{E}}$ is a real $3\times3$ matrix and $\mathbf{t}$ is
a translation vector. For unitary channels one has
$\mathbf{t}=0$ and $R_{\mathcal{E}}\in\mathrm{SO}(3)$, so that the
antisymmetric part
\begin{equation}
  A_{\mathcal{E}} =
  \frac{1}{2}\bigl(R_{\mathcal{E}} - R_{\mathcal{E}}^{\mathsf T}\bigr)
  = [\boldsymbol{\Omega}]_\times
\end{equation}
coincides with the antisymmetric generator $A_U$ of
Sec.~\ref{sec:quantum}. Therefore, once $R_{\mathcal{E}}$ has been
reconstructed from experimental data, the geometric content of a
unitary evolution is contained entirely in $A_{\mathcal{E}}$.

This observation suggests using the antisymmetric part of the
reconstructed adjoint map as a diagnostic tool for geometric phases
in noisy or imperfect implementations of qubit gates. In scenarios
where $\mathcal{E}$ is close to a unitary channel, $A_{\mathcal{E}}$
provides an effective geometric generator that can be compared with
the ideal design and used to detect deviations in the implemented
geometric phase. For example, in an experimental realization of a single-qubit geometric-phase gate in platforms such as superconducting circuits, trapped ions, or integrated photonics, one can reconstruct the process, extract the antisymmetric part of the adjoint map, and compare it directly with the ideal design to quantify geometric-phase errors even in the presence of moderate noise.

\subsection{Extensions beyond unitary two-level dynamics}

The present analysis focuses on unitary dynamics in two-level
systems, where the state space of pure states is the two-sphere and
the adjoint action of $\mathrm{SU}(2)$ is represented by
$\mathrm{SO}(3)$ rotations. It is natural to ask how far these
results can be extended.

One direction is to consider more general completely-positive
trace-preserving (CPTP) maps on qubits. In this case
$R_{\mathcal{E}}$ need not be orthogonal, and the antisymmetric
part $A_{\mathcal{E}}$ no longer generates a pure rotation.
Nevertheless, $A_{\mathcal{E}}$ continues to encode the infinitesimal
antisymmetric contribution to the mapping of Bloch vectors and may
serve as a useful indicator of geometric features even in the
presence of decoherence. When the channel is close to unitary, this antisymmetric contribution can be interpreted as an effective geometric generator, whereas deviations from orthogonality in $R_{\mathcal{E}}$ and the accompanying translation vector reflect the nonunitary corrections induced by noise and decoherence. A systematic analysis of geometric phases
for nonunitary evolutions in terms of $A_{\mathcal{E}}$ lies beyond
the scope of this work but appears as a natural extension.

Another direction is to explore higher-dimensional systems, such as
spin-$j$ particles or multilevel atoms, whose pure-state manifolds
can often be embedded in higher-dimensional analogues of the Bloch
sphere. In these cases the relevant Lie algebras are
$\mathfrak{su}(N)$ and their adjoint representations, which contain
antisymmetric generators associated with rotations in generalized
Bloch spaces. It is plausible that appropriate generalizations of the present construction will identify antisymmetric adjoint generators as carriers of geometric-phase information in those settings as well.

\section{Conclusions}
\label{sec:conclusions}

We have shown that the antisymmetric part of the adjoint
$\mathrm{SU}(2)$ generator provides a universal algebraic kernel for
geometric phases in two-level systems, encompassing both classical
polarization optics and quantum qubits. In the classical setting of
ideal retarders, the antisymmetric $3\times 3$ block of the Mueller
matrix is proportional to the cross-product matrix of a vector
$\boldsymbol{\Omega}$, which governs the angular velocity of the Stokes
vector on the Poincar\'e sphere. The Pancharatnam--Berry phase is
completely determined by the tangential component of this angular
velocity and can be expressed as $-\tfrac{1}{2}$ times the solid angle
enclosed by the trajectory of the Stokes vector. The symmetric part of
the Mueller block is geometrically neutral in the sense that it does
not contribute to the geometric phase.

In the quantum setting, the adjoint action of any
$\mathrm{SU}(2)$ unitary operator on the Bloch sphere is described by a
rotation matrix whose antisymmetric part is again of the form
$[\boldsymbol{\Omega}(t)]_\times$. The Bloch vector of a pure qubit
state evolves according to a kinematic equation driven by
$\boldsymbol{\Omega}(t)$, and the geometric phase associated with
cyclic evolutions is given by an expression formally identical to that
in the classical case. This holds independently of whether the
evolution is adiabatic or nonadiabatic, and it reproduces Berry’s
phase in the adiabatic limit. Geometrically, the antisymmetric
adjoint generator encodes not only the magnitude of the angular
velocity on the Poincar\'e/Bloch sphere but also its handedness, so
that it carries the chirality of the evolution and fixes the sign of
the geometric phase.

These results establish a direct and operational bridge between
classical polarization experiments, described by Mueller matrices, and
quantum two-level dynamics, described by unitary adjoint maps on the
Bloch sphere. In both cases, the complete geometric-phase content of the evolution is
encoded in the antisymmetric adjoint generator, which can be extracted
from measured Mueller matrices or from quantum process tomography. The present framework suggests several directions for future work. On
the classical side, the explicit identification of the antisymmetric
Mueller generator as the carrier of geometric-phase information
provides a natural tool to design and diagnose polarization-based
geometric-phase elements, including space-variant retarders,
geometric-phase lenses, metasurfaces, and birefringent fibers,
directly from their measured Mueller matrices. For space-variant devices such as q-plates (i.e., birefringent elements whose optical axis varies across the transverse plane) and related subwavelength-grating elements, the analysis can be applied locally (pixel by pixel) to the position-dependent Mueller/Jones description, so that the spatially varying antisymmetric generator directly encodes the designed geometric-phase wavefront~\cite{MarucciPRL2006,HasmanProgOpt2005}. On the quantum side,
the same antisymmetric generator can be exploited to analyze the
geometric content of single-qubit gates reconstructed via process
tomography and to explore extensions to nonunitary channels and
multilevel systems. In both regimes, the antisymmetric Mueller
generator offers a compact language to interpret experimental data in
terms of their underlying geometric structure.

\bigskip

\noindent\textbf{Funding.} No funding.

\smallskip
\noindent\textbf{Disclosures.} The author declares no conflicts of interest.

\smallskip
\noindent\textbf{Data availability.} No additional data were generated or analyzed in the course of this research.

\end{document}